\documentclass[journal]{IEEEtran}
\usepackage{amsmath,amsfonts}
\usepackage{algorithmic}
\usepackage{algorithm}
\usepackage{array}
\usepackage[caption=false,font=normalsize,labelfont=sf,textfont=sf]{subfig}
\usepackage{textcomp}
\usepackage{stfloats}
\usepackage{url}
\usepackage{verbatim}
\usepackage{graphicx}
\usepackage{cite}
\usepackage{hyperref}
\usepackage{booktabs}
\usepackage{xcolor}

\providecommand{\bm}[1]{\boldsymbol{#1}}

\begin{document}

\title{Agentic AI for Embodied-enhanced Beam Prediction in Low-Altitude Economy Networks

\author{Min Hao, Zhizhuo Li, Zirui Zhang, Maoqiang Wu, Han Zhang and Rong Yu}


\IEEEcompsocitemizethanks{
\IEEEcompsocthanksitem Min Hao, Zhizhuo Li, Maoqiang Wu and Han Zhang are with School of Electronic Science and Engineering, South China Normal University, Foshan, China (e-mail: haomin@scnu.edu.cn, 2025025406@m.scnu.edu.cn, maoqiangwu@m.scnu.edu.cn, zhanghan@scnu.edu.cn). 
\IEEEcompsocthanksitem Zirui Zhang is with School of Intelligent Engineering, Shaoguan university, Shaoguan, China (e-mail: ziruizhang9609@gmail.com).
\IEEEcompsocthanksitem Rong Yu is with School of Automation, Guangdong University of Technology, Guangzhou, China (e-mail: yurong@ieee.org).
\IEEEcompsocthanksitem \textit{Corresponding author: Zirui Zhang, Maoqiang Wu.}
}
}

\markboth{IEEE TRANSACTIONS ON COGNITIVE COMMUNICATIONS AND NETWORKING}%
{Shell \MakeLowercase{\textit{et al.}}: A Sample Article Using IEEEtran.cls for IEEE Journals}


\maketitle

\begin{abstract}
Millimeter-wave or terahertz communications can meet demands of low-altitude economy networks for high-throughput sensing and real-time decision making. However, high-frequency characteristics of wireless channels result in severe propagation loss and strong beam directivity, which make beam prediction challenging in highly mobile uncrewed aerial vehicles (UAV) scenarios. In this paper, we employ agentic AI to enable the transformation of mmWave base stations toward embodied intelligence. We innovatively design a multi-agent collaborative reasoning architecture for UAV-to-ground mmWave communications and propose a hybrid beam prediction model system based on bimodal data. The multi-agent architecture is designed to overcome the limited context window and weak controllability of large language model (LLM)-based reasoning by decomposing beam prediction into task analysis, solution planning, and completeness assessment. To align with the agentic reasoning process, a hybrid beam prediction model system is developed to process multimodal UAV data, including numeric mobility information and visual observations. The proposed hybrid model system integrates Mamba-based temporal modelling, convolutional visual encoding, and cross-attention-based multimodal fusion, and dynamically switches data-flow strategies under multi-agent guidance. Extensive simulations on a real UAV mmWave communication dataset demonstrate that proposed architecture and system achieve high prediction accuracy and robustness under diverse data conditions, with maximum top-1 accuracy reaching 96.57\%.
\end{abstract}


\section{Introduction}

\IEEEPARstart{T}{he} flourishing low-altitude economy is driving the emergence of various new industries, such as general aviation, low-altitude tourism, and low-altitude logistics,  among which uncrewed aerial vehicles (UAVs)  serve as the core enabling platforms\cite{11131292}. Meanwhile, the number of UAVs is rapidly growing worldwide. In particular, UAV density in urban areas is sharply increasing as low-altitude economy applications are being actively deployed. Enabling high-throughput and low-latency communication between a large number of UAVs and ground infrastructure is critical to ensuring the sustained and reliable development of the low-altitude economy. Millimeter-wave (mmWave) and terahertz (THz) frequency bands offer significantly larger bandwidth and higher spectrum reuse capability, and are widely regarded as promising frequency ranges for supporting high-density UAV deployments in 6G networks \cite{9598918,11087719,11130648}. However, the high mobility of UAVs and the instability of communication frequency bands pose significant challenges to UAV communications. 

Due to the severe path loss at mmWave and Terahertz frequencies, highly directional beamforming is indispensable for establishing reliable wireless links. To enable directional transmission, beam training is commonly adopted to establish bidirectional beam alignment between the transmitter and the receiver, especially during initial access and link recovery. In practical systems, beam training typically relies on beam search mechanisms, such as exhaustive or hierarchical beam sweeping, to identify suitable transmit–receive beam pairs. However, beam search incurs substantial training overhead and signalling latency, as multiple candidate beams need to be sequentially probed and evaluated. In UAV-to-ground communications, the overhead of beam search and beam training is further exacerbated by the three-dimensional mobility of UAVs, rapid trajectory variations, and frequent beam misalignment caused by attitude fluctuations and environmental dynamics. These challenges motivate a paradigm shift from reactive beam search–based training to proactive beam prediction, which aims to infer future beam directions in advance based on historical observations and contextual information. Nevertheless, conventional learning-based beam prediction methods often lack the ability to autonomously adapt to highly non-stationary air-to-ground channels \cite{10494372}. 

An agentic AI architecture built upon large language models (LLM) exhibits proactive perception and environment-aware reasoning capabilities, offering a promising solution for dynamic beam prediction in UAV communications. Unlike conventional beam prediction approaches that treat beam selection as a passive inference task, agentic AI enables beam prediction to be formulated as an embodied, goal-driven, and adaptive decision-making process. By explicitly modelling the perception–decision–action loop under high mobility and partial observability, agentic AI can proactively exploit multi-modal embodied information, optimize long-term communication objectives, and continuously adapt to dynamic low-altitude environments. This makes agentic AI a natural and powerful paradigm for next-generation beam prediction in low-altitude economy networks. However, when agentic AI is instantiated with large language models, limitations related to the context window and the lack of controllability in the reasoning process still remain. Addressing these LLM-centric challenges is critical to enabling reliable and practical agentic beam prediction in low-altitude economy networks\cite{11339915}.

In this study, we propose a multi-agent agentic AI architecture in which agents enable beam prediction for UAV mmWave communications through functional specialization. A complete workflow consisting of task understanding, tool invocation, reasoning and reflection, and solution deployment is accomplished through extensive inter-agent interactions. The data sources for beam prediction are multimodal. To align with the agentic AI architecture, we specifically design and train deep learning models tailored for multimodal data. The model framework and parameters can be adaptively adjusted based on the strategies provided by the agents, thereby improving prediction robustness in complex flight environments. The main contributions of this paper are as follows:

\begin{itemize}

\item To overcome the context window limitations of LLM-based agents and enhance the transparency of the reasoning process, we construct a multi-agent architecture consisting of a task analysis agent (TAA), a solution planning agent (SPA), and an completeness assessment agent (CAA). These agents are deployed at the ground base station and leverage the reasoning and acting (ReAct) framework to transform the base station into an embodied intelligent system capable of performing beam prediction.

\item To adapt to the proposed multi-agent reasoning architecture and efficiently process multimodal data, we propose a hybrid beam prediction model system. This hybrid model system leverages ResNet, Mamba, and Transformer to extract features from historical UAV flight numeric data, image data, and hybrid data, and to generate beam prediction results for UAV-to-ground base station communications. The proposed hybrid model system innovatively incorporates three types of triggers to enable flexible data feature extraction. The operating strategies of the triggers derived through interactive reasoning among multiple agents.

\item Numerical simulation experiments validate the effectiveness of the proposed architecture and models. We evaluate the format accuracy, content accuracy, semantic similarity and efficiency of the outputs generated by the three agents using different LLM backbones. We train and evaluate the hybrid model system using real-world UAV trajectory and beam prediction datasets. Experimental results show that the proposed hybrid model system achieves top-1 prediction accuracies of 84.13$\%$, 91.81$\%$ and 96.57$\%$ for numeric data, image data, and hybrid data, respectively. All experimental results demonstrate that the proposed multi-agent architecture and the hybrid model system are well aligned with each other and achieve high accuracy in beam prediction.

\end{itemize}

The remainder of this paper is organized as follows.
Section \ref{II} presents the related work. We describe the system models and agentic AI architecture in Section \ref{III}. In Section \ref{IV}, the details of the proposed multi-agent interaction algorithm and deep learning algorithms are presented. Section \ref{V} presents the simulation results, followed by the conclusion in Section \ref{vI}.

\section{Related Work}\label{II}
When UAVs communicate with ground base stations over mmWave links, beam training is a prerequisite for establishing reliable connections. However, this process faces a series of structural challenges in low-altitude, highly mobile scenarios. This section systematically reviews existing approaches and highlights their limitations.

\subsection{Traditional Beam Training for UAV Communications}
UAV-to-ground mmWave communications rely on highly directional beamforming to overcome severe path loss. However, since the optimal transmit–receive beam directions are unknown prior to link establishment and vary over time due to UAV mobility, beam training is required to identify and align the appropriate beam pair before data transmission. 

In \cite{8056991}, the authors propose pseudo-exhaustive beam training and probabilistic beam tracking, achieving a favourable trade-off among training overhead, robustness, and throughput. In \cite{9880856}, the UAV angular velocity is estimated during initial training to determine beam coherence time, enabling beam tracking with only four candidate beams per update and significantly reduced overhead while preserving high uplink capacity. For UAV mmWave communications with airframe blockage, \cite{10256089} employs cylindrical conformal antenna arrays to achieve full-space coverage and designs a hierarchical codebook with a corresponding search algorithm for efficient beam training under blockage-induced attenuation. Further, bidirectional angle-aware beam tracking and adaptive beam reconstruction are proposed in \cite{10505822}. An improved global dynamic crow search algorithm enables fast localization and angle tracking at the UAV base station without historical information, while Gaussian process regression predicts user orientation to reduce pilot overhead, jointly supporting timely beam reconstruction and improved SNR in high-mobility scenarios. In \cite{11176943}, the authors construct a joint communication-and-sensing near-field signal model to address frequent beam misalignment in UAV communications, and implement joint beam tracking based on an extended Kalman filter. This work provides a new perspective on low-overhead and robust beam alignment for highly mobile UAVs in 6G near-field communications.

The above methods mainly aim to reduce beam training overhead, since conventional beam training and search schemes are insufficiently robust to UAV high mobility and dynamic environments. Existing solutions largely rely on passive adaptation to mobility, lacking intelligent and proactive beam decision mechanisms.

\subsection{Learning-based Beam Prediction Methods}
Traditional beam training relies on exhaustive or hierarchical beam search, incurring high overhead and becoming inefficient in high-mobility UAV scenarios. In contrast, learning-based beam prediction leverages historical and contextual information to infer future beam directions, reducing training latency and improving robustness.

To address beam misalignment caused by airframe jitter in UAV communications, the authors propose a learning-driven predictive beamforming approach\cite{9143143}. The proposed method employs a long short-term memory (LSTM) recurrent neural network to predict pitch-related angles in the next time slot from historical angle sequences, thereby enabling fast tracking and robust alignment of jitter-induced angular fluctuations. In \cite{9813590}, a dual-band cooperative access architecture integrating sub-6 GHz and mmWave is proposed for UAV-assisted high-speed railway communications. To combat frequent beam misalignment under dual high mobility, Kalman filtering combined with LSTM is used for beam angle prediction and tracking, significantly reducing training and feedback overhead while improving spectral efficiency and transmission reliability. In \cite{11173430}, a multimodal sensing–assisted deep learning framework is proposed, in which base-station RGB images and UAV location information are fused to directly predict the optimal beam without beam training, and a real-world field dataset is constructed and released for synchronized evaluation. In \cite{10912462}, a multimodal learning–based beam prediction method is proposed, in which image and mmWave radar information are fused via cross-modal feature enhancement and an uncertainty-aware dynamic fusion mechanism to achieve robust optimal beam prediction.

The above methods rely on deep learning for beam prediction. Existing learning-based beam prediction methods typically treat beam selection as a static inference problem, lacking autonomous adaptation and proactive decision-making under highly dynamic UAV environments.
 
\subsection{LLM-based Intelligent Communication Systems}
LLM-based beam prediction enables reasoning-driven, context-aware, and adaptive beam decision-making, which goes beyond static inference and significantly improves robustness under highly dynamic UAV communication environments.

In \cite{11308125}, the challenges of mmWave beam prediction under complex weather and traffic conditions are addressed by introducing a pretrained LLM as a reasoning engine with efficient cross-scenario adaptation and knowledge transfer mechanisms. In \cite{10892257}, the authors leverage a LLM as a general-purpose temporal reasoning engine and transform the wireless beam evolution problem into an LLM-understandable prediction task through sequence reprogramming and prompt learning. The LLM-based approach consistently outperforms LSTM-based baselines in terms of robustness across multiple aspects. In \cite{11313491}, the authors address the beam prediction challenge in near and far field mixed, non-stationary, time-varying channels of mmWave massive MIMO under high-mobility scenarios, and propose a BP-GPT framework based on a pretrained generative large language model (GPT-2). In \cite{11157846}, the authors propose a UAV communication optimization framework integrating a multimodal LLM with deep reinforcement learning. The LLM fuses perception and semantic information to predict user trajectories and enable joint trajectory and beamforming optimization.

The above methods rely on a single LLM for beam prediction. In contrast, a multi-agent architecture enables structured reasoning, strategy-level adaptation, and coordinated use of heterogeneous prediction tools, which is crucial for reliable beam management in highly dynamic UAV mmWave communications. To the best of our knowledge, few studies have explored multi-agent LLM-based agentic architectures for embodied beam prediction in low-altitude UAV scenarios.

\section{System Model and Architecture}\label{III}
\begin{figure*}[!t]
\centering
\includegraphics[width=1.0\textwidth]{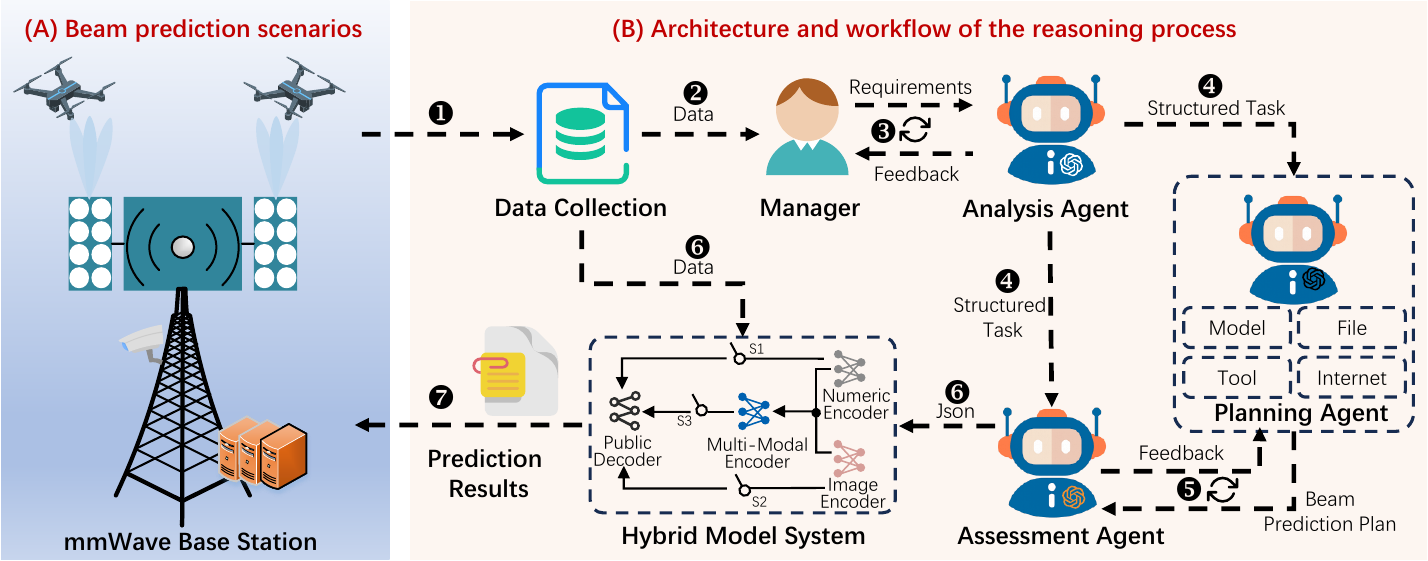} 
    \caption{UAV-to-ground mmWave beam prediction scenarios and reasoning architecture. \textit{Part A} illustrates the beam prediction scenario, which includes a mmWave base station equipped with high-definition cameras and UAVs equipped with beam transceivers.\textit{Part B} presents the architecture and workflow of joint reasoning and beam prediction through the cooperation of the multi-agent and the hybrid model. Inter-agent interactions and reasoning are carried out sequentially following the numbered steps.}  
    \label{fig1}         
\end{figure*}

This section first presents a mathematical formulation of the beam prediction problem, followed by a description of the overall architecture and workflow of embodied beam prediction.

\subsection{System Model for Beam Prediction}
The scenario of UAV–to–ground base station communication based on mmWave transmission and beam prediction is illustrated in Part A of Fig.~\ref{fig1}. The mmWave base station is equipped with a uniform linear array (ULA) consisting of $M$ elements and a camera that captures RGB images of the UAV. The base station adjusts the beamforming parameters based on the UAV's GPS location and image data, thereby enhancing the SNR between the UAV and the base station. The base station possesses a predefined codebook for mmWave beamforming, denoted as ${\cal B} = \left\{ {{b_q}} \right\}_{q = 1}^Q$. Total of $Q$ beamforming vectors are stored in the codebook, and each beamforming vector is a complex-valued vector ${b_q} \in \mathbb{C}^{M \times 1}$. In a beamforming codebook, each beamforming vector is indexed by a beam index. The mmWave communication system employs an orthogonal frequency division multiplexing (OFDM) scheme, which divides a high-speed data stream into $K$ low-speed subcarriers and uses a cyclic prefix of length $D$. When the UAV flies over the base station at time $t$, the signal carried by the $k$-th subcarrier can be expressed as:
\begin{equation}\label{eq:sys_form_1}
  {y_k}\left[ t \right] = {h_k}\left[ t \right]{b_q}\left[ t \right]x + {\beta _k}\left[ t \right],
\end{equation}
where ${h_k}\left[ t \right]$ denotes the channel between the UAV and the base station at time $t$, and $x$ represents the signal transmitted by the UAV. The ${\beta _k}\left[ t \right]$ denotes complex Gaussian noise, which follows the distribution $\mathcal{N}_{\mathbb{C}}(0, \sigma^2)$. All variables in formula (\ref{eq:sys_form_1}) belong to the complex domain. The channel ${h_k}\left[ t \right]$ at time $t$ can be calculated as follows:
\begin{equation}\label{eq:sys_form_2}
{h_k}\left[ t \right] = \sum\limits_{d = 0}^{D - 1} {\sum\limits_{n = 1}^N {{\alpha _n}} } {e^{ - j\frac{{2\pi k}}{K}d}}P\left( {d\Delta t - {\tau _n}} \right)\left( {{\theta _n},{\phi _n}} \right),
\end{equation}
where $N$ denotes the number of channel paths. ${{\alpha _n}}$, ${{\tau _n}}$, ${{\theta _n}}$, and ${{\phi _n}}$ represent the path loss, time delay, azimuth angle of arrival, and elevation angle of arrival for the $n$-th channel path, respectively. $P$ denotes the transmission power, and ${\Delta t}$ represents the sampling time.

Beam training aims to select the optimal ${{b_q}}$ from the codebook at the current time to maximize the SNR between the UAV and the base station. This can be mathematically formulated as:
\begin{equation}\label{eq:sys_form_3}
\widehat b\left[ t \right] = \mathop {\arg \max }\limits_{{b_q}\left[ t \right] \in {\cal B}} \frac{1}{K}\sum\limits_{k = 1}^K {SNR{{\left| {{h_k}\left[ t \right]{b_q}\left[ t \right]} \right|}^2}}.
\end{equation}
In this study, it is assumed that the channel remains time-invariant within each channel coherence interval. The channel state is updated only when the system enters the next time block.

Ideally, the mmWave base station should be able to select the optimal beamforming vector $\widehat b\left[ t \right]$ from the codebook at any time 
$t$. A direct approach for the mmWave base station to obtain the optimal beamforming vector is to compute it based on explicit channel knowledge. However, the strong environmental sensitivity of mmWave signals and the high mobility of UAVs make accurate acquisition of explicit channel knowledge extremely challenging. Alternatively, the mmWave base station can perform various forms of codebook search, but this typically incurs prohibitively high beam training overhead. In this paper, we define the mmWave base station as an embodied intelligent base station which capable of sensing, reasoning, and acting. It performs deep learning–based beam prediction using the collected UAV-related data. The mmWave base station is equipped with a high-definition camera, which captures video and image data of UAV flights. In addition, numeric data such as the UAV’s GPS information, altitude, velocity, and its distance to the base station are transmitted back to the base station. We define UAV flight images as:
\begin{equation}
\mathbf{I}_t \in \mathbb{R}^{C \times H \times W},\quad t = 1, \ldots, T.
\end{equation}
Numeric data of UAV flight are represented as:
\begin{equation}
\mathbf{x}_t \in \mathbb{R}^{d_n},
\quad t = 1, \ldots, T ,
\end{equation}
where ${d_n}$ denotes dimension of numeric data. Accordingly, the data perceived by the mmWave base station over the time interval $\tau $ are given by:
\begin{equation}
{S_t} = \left\{ {{{\bf{I}}_t},{{\bf{x}}_t}} \right\},\quad t = 1, \ldots ,T.
\end{equation}

Beam prediction aims to establish a mapping between the sensed data ${S_t}$ at the mmWave base station and the optimal beamforming vectors in a future time interval, with the mapping function serving as the beam prediction model. The mapping function can be formally expressed as:
\begin{equation}
{f_\theta }:{S_t} \to \hat b\left[ t \right].
\end{equation}
Under multi-agent collaborative reasoning, the mmWave base station employs a hybrid deep learning model system for beam prediction. Then, the objective of the hybrid model system is to maximize the number of correctly predicted samples in ${S_t}$. Accordingly, the problem can be formally formulated as:
\begin{equation}
{f_{{\theta ^*}}}^* = \mathop {\arg \max }\limits_\theta  \prod\limits_{u = 1}^U {\left( {{{\hat b}_u}\left[ t \right] = {b_q}\left[ t \right]\left| {{S_t}} \right.} \right)} ,
\end{equation}
where $\theta $ denotes the parameters of the hybrid model system and $U$ is the
total number of samples in the dataset

\subsection{Architecture and Workflow}
Part B of Fig.~\ref{fig1} illustrates the embodied reasoning architecture of the mmWave base station, which mainly consists of three interacting LLM agents and a hybrid model system. As discussed above, the image data and numeric data from the UAV are collected by the mmWave base station, after which the base station manager formulates the beam prediction requirements based on the available data.
\subsubsection{Task Analysis Agent}
The TAA serves as reasoning module for manager’s requirements. It receives the beam prediction requirements from the manager in natural language form and transforms them into structured task specifications. The TAA focuses on key elements of the manager’s requirements, including the number of UAVs, data storage locations, data formats, data labels, and prediction accuracy requirements. If any of this critical information is missing, the agent continues to interact with the manager to obtain the missing details. Meanwhile, the TAA rewrites the manager’s requirements using more concise and precise language. The rewritten requirements, together with the key information, form a structured task specification that is transmitted to the other agents.

\subsubsection{Solution Planning Agent}
The SPA generates a beam prediction plan based on the structured tasks produced by TAA. The SPA can invoke pre-trained beam prediction models along with their parameter files and can also employ external tools to assess data quality. For example, when the data include both image and numeric modalities, the SPA can employ an image blurriness assessment tool. If the image blurriness exceeds a predefined threshold, only the numeric data are used as the sole input to the beam prediction model to improve prediction accuracy. The SPA can also perform retrieval based on retrieval-augmented generation (RAG) over Internet and external document repositories, such as relevant technical papers and communication standards, thereby improving the reliability of the generated plan. 

Unlike conventional LLM reasoning, the SPA does not generate a plan through single-step inference but follows a ReAct-based reasoning paradigm. ReAct reasoning enables LLM to interleave explicit reasoning with action execution, allowing them to interact with external tools and environments in a closed-loop manner. We can configure the number of reasoning iterations performed by SPA before producing a final result, or determine the reasoning iterations by setting a semantic similarity threshold between the final result and a reference answer.

The generated beam prediction plan mainly includes a data-flow switching strategy for the hybrid model system and a parameter configuration for the selected prediction model. The switching strategy is determined by input data types, while the parameter configuration of the prediction model is determined by prediction accuracy requirements and computational capability of mmWave base station. After the overall beam prediction plan is generated, it is transmitted to CAA for evaluation.

\subsubsection{Completeness Assessment Agent}
The function of CAA is to assess the completeness of the alignment between the plan generated by SPA and the tasks specified by TAA. When the plan satisfies task requirements, CAA outputs task status as resolved and generates JSON-formatted commands based on the plan to activate hybrid model system. When the plan does not satisfy task requirements, CAA outputs task status as unresolved and provides reasoning-based explanations for the unresolved causes. The identified causes are sent to SPA, and the two agents interact iteratively until the task is resolved.

\subsubsection{Hybrid Model System}
The hybrid model system includes three encoders and a public decoder. The encoders extract spatio-temporal features from input data, where numeric and image encoders process UAV data sensed by the mmWave base station, and a multimodal encoder fuses features from numeric and image data. The public decoder takes data features as input and outputs beam prediction results. The hybrid model system contains three data-flow switches corresponding to three beam prediction schemes, namely prediction based on numeric data features, prediction based on image data features, and prediction based on hybrid data features. The data-flow switching strategy is obtained through the interactive reasoning among the aforementioned agents.

\section{Algorithm Design}\label{IV}
This section details algorithms and training methods involved in multi-agent architecture and hybrid model system.

\subsection{Multi-Agent Algorithmic Workflow}
\begin{algorithm}[!t]
\caption{Multi-Agent Requirement-to-Plan Workflow}
\label{alg:multi_agent_workflow}
\begin{algorithmic}[1]
\REQUIRE Natural-language requirement $R$
\ENSURE Final plan $S$

\STATE $isTaskClear \leftarrow \textbf{false}$
\STATE $isPlanValid \leftarrow \textbf{false}$
\STATE $T \leftarrow \emptyset$ \COMMENT{Structured task}
\STATE $S \leftarrow \emptyset$ \COMMENT{Candidate plan}
\STATE $F \leftarrow \emptyset$ \COMMENT{Feedback from CAA}
\STATE $c \leftarrow 0$, $C_{\max} \leftarrow C$ \COMMENT{Max validation iterations}

\STATE \COMMENT{\textbf{TAA: rewrite requirement as a clear task; ask questions if needed}}
\WHILE{$isTaskClear = \textbf{false}$}
    \STATE $T \leftarrow \textsc{TAA\_Rewrite}(R)$
    \IF{$\textsc{TAA\_IsClear}(T) = \textbf{false}$}
        \STATE $Q \leftarrow \textsc{TAA\_Ask}(T)$
        \STATE $R \leftarrow \textsc{GetManagerResponse}(Q)$
    \ELSE
        \STATE $isTaskClear \leftarrow \textbf{true}$
    \ENDIF
\ENDWHILE

\STATE \COMMENT{\textbf{SPA \& CAA: plan generation and iterative completeness assessment}}
\WHILE{$isPlanValid = \textbf{false} \;\AND\; c < C_{\max}$}
    \STATE $S \leftarrow \textsc{SPA\_Generate}(T, F)$
    \STATE $(isPlanValid, F) \leftarrow \textsc{CAA\_Evaluate}(S, T)$
    \STATE $c \leftarrow c + 1$
\ENDWHILE

\end{algorithmic}
\end{algorithm}

Algorithm \ref{alg:multi_agent_workflow} presents a multi-agent requirement-to-plan workflow that transforms a user’s natural-language requirement into a validated execution plan through structured task understanding and iterative refinement. 

Specifically, the algorithm first initializes a set of control flags and intermediate variables to track task clarity, plan validity, structured task representation, candidate plans, and evaluator feedback (Lines 1–6). Given an initial natural-language requirement $R$, the TAA rewrites the requirement into a structured task description $T$ and checks whether the task is sufficiently clear (Lines 7–16). If ambiguities remain, the TAA actively generates clarification questions and updates the requirement based on manager responses, ensuring that subsequent planning is grounded in a well-defined task.

Once a clear task representation is obtained, the algorithm enters an iterative plan generation and assessment phase jointly handled by the SPA and CAA (Lines 17–21). The SPA generates a candidate plan $S$ conditioned on the structured task $T$ and the feedback $F$ from previous evaluation rounds. The CAA then evaluates whether the generated plan satisfies the task requirements and provides explicit feedback when deficiencies are identified. This feedback-driven loop continues until the plan is deemed valid or a maximum number of validation iterations is reached, which guarantees termination.

This algorithm explicitly separates task understanding, plan generation, and plan assessment into dedicated agent roles, enabling systematic reasoning, error diagnosis, and iterative improvement. Such a design is particularly suitable for complex planning scenarios where requirements are initially underspecified and plan correctness must be assessed against task-level constraints rather than surface-level outputs.

All three agents in Algorithm \ref{alg:multi_agent_workflow} follow ReAct paradigm for reasoning. Each ReAct step includes reasoning, action, and observation. Let $x$ denote agent prompt, ${r_t}$ denote reasoning text at step $t$, ${a_t}$ denote action at step $t$, and ${o_t}$ denote observation obtained after executing the action. Up to step $t$, reasoning history or context can be constructed as:
\begin{equation}
h_t = \left( x,\; r_{1:t-1},\; a_{1:t-1},\; o_{2:t} \right).
\end{equation}
Core idea of ReAct is to generate reasoning and actions conditioned on historical context, which can be expressed as:
\begin{equation}
\pi_\theta(r_t, a_t \mid h_t)
=
\pi_\theta(r_t \mid h_t)\,
\pi_\theta(a_t \mid h_t, r_t),
\end{equation}
where $\theta $ denotes parameters of LLM used by agent. ReAct can be viewed as a two-stage process, where reasoning ${r_t}$ is first generated and action ${a_t}$ is then selected under guidance of this reasoning. Action execution yields observation ${o_{t + 1}}$, which is then used to update reasoning history:
\begin{equation}
h_{t+1}=h_t \cup \{ r_t,\; a_t,\; o_{t+1} \}.
\end{equation}

Overall reasoning process terminates when a predefined objective is achieved, such as TAA obtaining all key beam prediction requirement information, SPA generating a beam prediction plan that reaches a semantic similarity threshold, and CAA determining that beam prediction plan provides a point-to-point response to task requirements.

\begin{algorithm}[t]
\caption{Hybrid Beam Prediction Model System}
\label{algor_hybrid}
\begin{algorithmic}[1]
\REQUIRE Numeric sequence $\bm{X}$, image sequence $\bm{I}$, inference mode $\mathcal{M}\in\{\mathtt{Numeric},\mathtt{Image},\mathtt{Multi}\}$
\ENSURE Predicted beam index sequence $\hat{\bm{B}}$

\IF{$\mathcal{M}=\mathtt{Numeric}$}
    \STATE \textbf{Enable} switches $S_{11}$ and $S_{12}$
    \STATE Obtain numeric embeddings: $\bm{X}_{emb} \leftarrow \mathrm{PE}(\mathrm{Proj}(\bm{X}))$
    \STATE Extract numeric temporal features: $\bm{X}_{tmp} \leftarrow \mathrm{Mamba2}^{(N)}(\bm{X}_{emb})$
    \STATE Decode to prediction: $\hat{\bm{B}} \leftarrow \mathrm{Dec}(\bm{X}_{tmp})$
\ELSIF{$\mathcal{M}=\mathtt{Image}$}
    \STATE \textbf{Enable} switches $S_{21}$ and $S_{22}$
    \STATE Obtain image embeddings: $\bm{I}_{emb} \leftarrow \mathrm{PE}(\mathrm{Proj}(\bm{I}))$
    \STATE Extract image temporal features: $\bm{I}_{tmp} \leftarrow \mathrm{Mamba2}^{(N)}(\bm{I}_{emb})$
    \STATE Decode to prediction: $\hat{\bm{B}} \leftarrow \mathrm{Dec}(\bm{I}_{tmp})$
\ELSE
    \STATE \textbf{Enable} switches $S_{11}$, $S_{21}$, and $S_{3}$ \hfill \textit{(Multi)}
    \STATE Obtain numeric embeddings: $\bm{X}_{emb} \leftarrow \mathrm{PE}(\mathrm{Proj}(\bm{X}))$
    \STATE Extract numeric temporal features: $\bm{X}_{tmp} \leftarrow \mathrm{Mamba2}^{(N)}(\bm{X}_{emb})$
    \STATE Obtain image embeddings: $\bm{I}_{emb} \leftarrow \mathrm{PE}(\mathrm{Proj}(\bm{I}))$
    \STATE Extract image temporal features: $\bm{I}_{tmp} \leftarrow \mathrm{Mamba2}^{(N)}(\bm{I}_{emb})$
    \STATE Fuse modalities: $\bm{M} \leftarrow \mathrm{MMEnc}(\bm{X}_{tmp}, \bm{I}_{tmp})$
    \STATE Decode to prediction: $\hat{\bm{B}} \leftarrow \mathrm{Dec}(\bm{M})$
\ENDIF

\STATE \textbf{return} $\hat{\bm{B}}$
\end{algorithmic}
\end{algorithm}

\subsection{Hybrid Model System Algorithmic Workflow}
\begin{figure*}[!t]
\centering
\includegraphics[width=1.0\textwidth]{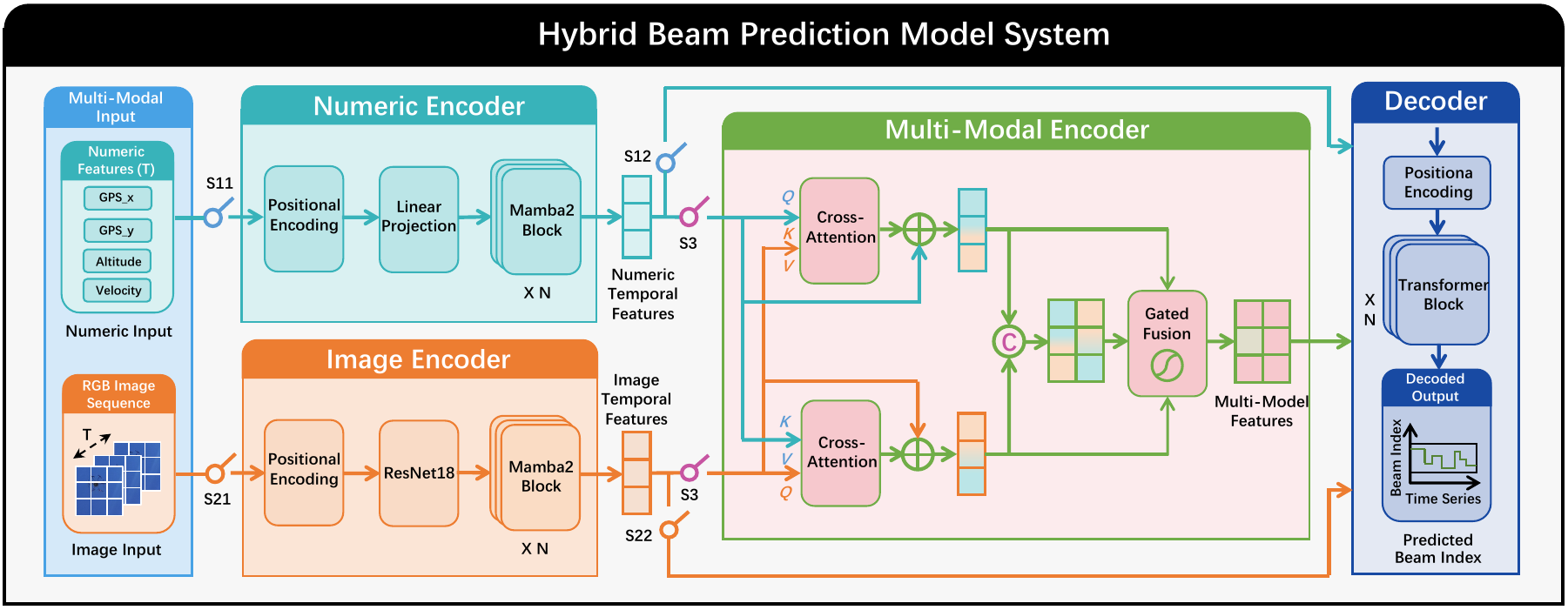} 
    \caption{Components and Workflow of Hybrid Beam Prediction Model System. Numeric and image data of UAVs sensed by mmWave base station are processed by encoders for feature extraction and then passed to a decoder to obtain beam prediction results. Specific data-flow switching strategies and encoder parameters are obtained through multi-agent interactive reasoning.}  
    \label{fig_hybrid}         
\end{figure*}

The proposed hybrid beam prediction model system integrates heterogeneous numeric and visual modalities to enable robust beam index prediction under highly dynamic UAV communication scenarios. As illustrated in Fig.~\ref{fig_hybrid}, the system consists of a multi-modal input module, modality-specific encoders, a multi-modal interaction encoder, and a sequence decoder\cite{liu2025timecma,wang2025chattime,huang2025causal}.

Specifically, the numeric branch processes time-series mobility-related features (e.g., GPS coordinates, altitude, and velocity) through positional encoding and linear projection, followed by stacked Mamba blocks to capture long-range temporal dependencies in numeric signals. In parallel, the visual branch extracts spatio-temporal representations from RGB image sequences using positional encoding, a ResNet-based backbone, and Mamba blocks to model temporal correlations across consecutive frames. These two encoders generate modality-specific temporal feature embeddings with aligned sequence lengths.

To effectively exploit cross-modal complementarities, a multi-modal encoder is designed based on bidirectional cross-attention mechanisms, where numeric features attend to visual features and vice versa. The resulting representations are aggregated via residual connections and concatenation, and further refined through a gated fusion module, which adaptively controls the contribution of each modality according to the current context. This design enables dynamic modality weighting and enhances robustness against modality degradation or uncertainty.

Finally, fused multimodal features are fed into a Transformer-based positional encoding decoder to model temporal evolution and output beam index sequences. Under control of data-flow switching, decoder can also perform beam prediction using only features from numeric data or image data. By jointly exploiting numeric motion cues and visual environmental information, proposed hybrid architecture achieves accurate and stable beam prediction while maintaining flexible multimodal extensibility.

We provide a detailed breakdown of all encoders, decoder, and model training process in the following. Observation window length is defined as $T$, and prediction horizon is defined as $P$. Objective of hybrid model system is to output beam labels for next $P$ steps:
\begin{equation}
    y_\tau \in \{1, \ldots, K\},
\quad \tau = 1, \ldots, P .
\end{equation}

\subsubsection{Numeric Encoder}
Given a numeric feature sequence with length $T$:
\begin{equation}
    \mathbf{X}
=
\left[
\mathbf{x}_1,\,
\mathbf{x}_2,\,
\ldots,\,
\mathbf{x}_T
\right]^{\top}
\in
\mathbb{R}^{T \times d_n},
\end{equation}
where $\mathbf{x}_t \in \mathbb{R}^{d_n}
$ represents numeric feature vector at time step $t$, and ${d_n}$ denotes numeric feature dimension. Objective of numeric encoder is to map $ \mathbf{X}$ into latent space, which is expressed as:
\begin{equation}
    \mathbf{H}^n
=
\left[
\mathbf{h}_1^{n},\,
\mathbf{h}_2^{n},\,
\ldots,\,
\mathbf{h}_T^{n}
\right]^{\top}
\in
\mathbb{R}^{T \times d},
\end{equation}
where $d$ denotes model hidden dimension.

First, input features are linearly projected to unify representation dimensions across modalities, which is expressed as:
\begin{equation}
    \mathbf{E}_t
=
\mathbf{W}_n \mathbf{x}_t
+
\mathbf{b}_n,
\quad
t = 1, \ldots, T.
\end{equation}
All time steps are stacked to obtain ${\bf{E}} = {\left[ {{{\bf{E}}_1},{\kern 1pt}  \ldots ,{\kern 1pt} {{\bf{E}}_T}} \right]^ \top }$.

Then, positional encoding $\mathbf{P} \in \mathbb{R}^{T \times d}$ is introduced to explicitly inject temporal order information and added to projected features, which is expressed as:
\begin{equation}
    \mathbf{H}^{(0)} = \mathbf{E} + \mathbf{P},
\end{equation}
where $\mathbf{H}^{(0)} \in \mathbb{R}^{T \times d}$ denotes initial hidden representation before entering stacking module.

Subsequently, $N$ Mamba residual blocks are stacked to model long-range temporal dependencies. A Pre-LayerNorm structure is adopted in the $l$-th residual block, which is expressed as:
\begin{equation}
    {X_{emb}} = {\widehat {\bf{H}}^{(l - 1)}} = {\rm{LN}}\left( {{{\bf{H}}^{(l - 1)}}} \right),l = 1, \ldots ,N.
\end{equation}
Normalized representation is fed into Mamba module to extract temporal features, which is expressed as:
\begin{equation}
{{\bf{M}}^{(l)}} = {\mathop{\rm Mamba}\nolimits} \left( {{{\widehat {\bf{H}}}^{(l - 1)}}} \right).
\end{equation}
Output of layer $l$ is obtained through residual connection:
\begin{equation}
    \mathbf{H}^{(l)}
=
\mathbf{H}^{(l-1)}
+
\operatorname{Dropout}\!\left(
\mathbf{M}^{(l)}
\right).
\end{equation}
The above recursion proceeds from $l=1$ to $l=N$, yielding the final output of the numeric encoder as:
\begin{equation}
    \mathbf{H}^n = \mathbf{H}^{(N)} \in \mathbb{R}^{T \times d}.
\end{equation}

\subsubsection{Image Encoder}
Similar to the numeric encoder, given an RGB image sequence with length $T$:
\begin{equation}
    \mathbf{I}
=
\{ \mathbf{I}_1,\,
\mathbf{I}_2,\,
\ldots,\,
\mathbf{I}_T \},
\qquad
\mathbf{I}_t \in \mathbb{R}^{C \times H \times W}.
\end{equation}
Objective of image encoder is to map image sequence into a visual temporal latent representation:
\begin{equation}
    \mathbf{H}^v
=
\left[
\mathbf{h}_1^{v},\,
\mathbf{h}_2^{v},\,
\ldots,\,
\mathbf{h}_T^{v}
\right]^{\top}
\in
\mathbb{R}^{T \times d}.
\end{equation}

First, a convolutional backbone network is used to extract global semantic features from each image frame and stack them, which are respectively represented as:
\begin{equation}
    \mathbf{f}_t
=
\operatorname{ResNet18}\!\left(
\mathbf{I}_t
\right)
\in
\mathbb{R}^{d_r},
\quad
t = 1, \ldots, T ,
\end{equation}
\begin{equation}
    \mathbf{F}
=
\left[
\mathbf{f}_1,\,
\ldots,\,
\mathbf{f}_T
\right]^{\top}
\in
\mathbb{R}^{T \times d_r}.
\end{equation}

Then, to enable interaction with numeric modality in same latent space, visual features are projected to $d$-dimensional space and stacked, which is expressed as:
\begin{equation}
    \mathbf{E}_t^{v}
=
\mathbf{W}_v \mathbf{f}_t
+
\mathbf{b}_v,
\quad
\mathbf{W}_v \in \mathbb{R}^{d \times d_r},
\;
\mathbf{b}_v \in \mathbb{R}^{d},
\end{equation}
\begin{equation}
    \mathbf{E}^{v}
=
\left[
\mathbf{E}_1^{v},\,
\ldots,\,
\mathbf{E}_T^{v}
\right]^{\top}
\in
\mathbb{R}^{T \times d}.
\end{equation}
To explicitly inject temporal order information, positional encoding $\mathbf{P} \in \mathbb{R}^{T \times d}$ is introduced to obtain initial representation:
\begin{equation}
    \mathbf{G}^{(0)} = \mathbf{E}^{v} + \mathbf{P}.
\end{equation}

Finally, $N$ Mamba residual blocks are stacked to model inter-frame temporal dependencies. A Pre-LayerNorm structure is adopted in the $l$-th residual block, which is expressed as:
\begin{equation}
\left\{ {\begin{array}{*{20}{l}}
{{I_{emb}} = {{{\bf{\tilde G}}}^{(l - 1)}} = {\rm{LN}}\left( {{{\bf{G}}^{(l - 1)}}} \right)}\\
{{{\bf{Q}}^{(l)}} = {\rm{Mamba}}\left( {{{{\bf{\tilde G}}}^{(l - 1)}}} \right)}\\
{{{\bf{G}}^{(l)}} = {{\bf{G}}^{(l - 1)}} + {\rm{Dropout}}\left( {{{\bf{Q}}^{(l)}}} \right)}
\end{array}} \right.,
\end{equation}
and final output $\mathbf{H}^{v} = \mathbf{G}^{(N)} \in \mathbb{R}^{T \times d}$.

\subsubsection{Multi-Modal Encoder}
Given numeric modality latent representation $\mathbf{H}^{n}$ and visual modality latent representation $\mathbf{H}^{v}$, objective of multi-modal encoder is to explicitly model cross-modality dependencies and obtain fused memory $\mathbf{M} \in \mathbb{R}^{T \times d}$.

First, standard scaled dot-product attention is defined as:
\begin{equation}
    \operatorname{Attn}(\mathbf{Q}, \mathbf{K}, \mathbf{V})
=
\operatorname{softmax}\!\left(
\frac{\mathbf{Q}\mathbf{K}^{\top}}{\sqrt{d}}
\right)
\mathbf{V}.
\end{equation}
Multi-modal encoder adopts Multi-Head Attention (MHA), which can be written as:
\begin{equation}
    \left\{ \begin{array}{l}
{\mathop{\rm MHA}\nolimits} ({\bf{Q}},{\bf{K}},{\bf{V}}) = {\mathop{\rm Concat}\nolimits} \left( {{\rm{hea}}{{\rm{d}}_1},{\kern 1pt}  \ldots ,{\kern 1pt} {\rm{hea}}{{\rm{d}}_h}} \right){{\bf{W}}_O}\\
{\rm{hea}}{{\rm{d}}_i} = {\mathop{\rm Attn}\nolimits} \left( {{\bf{QW}}_i^Q,{\kern 1pt} {\bf{KW}}_i^K,{\kern 1pt} {\bf{VW}}_i^V} \right)
\end{array} \right.,
\end{equation}

Second, visual representations are used as Query, while numeric representations serve as Key and Value, thereby injecting numeric information into visual representations:
\begin{equation}
    \mathbf{U}^{v}
=
\operatorname{LN}
\left(
\mathbf{H}^{v}
+
\operatorname{MHA}
\left(
\mathbf{H}^{v},\,
\mathbf{H}^{n},\,
\mathbf{H}^{n}
\right)
\right),
\end{equation}
and A feed-forward network (FFN) is applied for non-linear transformation and residual formation:
\begin{equation}
    \mathbf{Z}^{v}
=
\operatorname{LN}
\left(
\mathbf{U}^{v}
+
\operatorname{FFN}
\left(
\mathbf{U}^{v}
\right)
\right).
\end{equation}

Third, symmetrically, numeric representations are used as Query, while visual representations serve as Key and Value to inject visual information into numeric representations, which is expressed as:
\begin{equation}
    \mathbf{U}^{n}
=
\operatorname{LN}
\left(
\mathbf{H}^{n}
+
\operatorname{MHA}
\left(
\mathbf{H}^{n},\,
\mathbf{H}^{v},\,
\mathbf{H}^{v}
\right)
\right),
\end{equation}
\begin{equation}
    \mathbf{Z}^{n}
=
\operatorname{LN}
\left(
\mathbf{U}^{n}
+
\operatorname{FFN}
\left(
\mathbf{U}^{n}
\right)
\right).
\end{equation}

Finally, to adaptively control contributions of two modalities under different scenarios, a gating vector $\mathbf{G} \in \mathbb{R}^{T \times d}.
$ is introduced:
\begin{equation}
    \mathbf{G}
=
\sigma
\left(
\mathbf{W}_g
\left[
\mathbf{Z}^{v} \,\|\, \mathbf{Z}^{n}
\right]
+
\mathbf{b}_g
\right),
\end{equation}
where ${{\bf{Z}}^v}\left\| {{{\bf{Z}}^n}} \right.$ denotes feature concatenation and $\sigma \left(  \right)$ denotes sigmoid function. Final fused representation is defined by element-wise weighting:
\begin{equation}
    \mathbf{M}
=
\mathbf{G} \odot \mathbf{Z}^{v}
+
\left(
\mathbf{1} - \mathbf{G}
\right)
\odot
\mathbf{Z}^{n}.
\end{equation}
This fusion form enables dynamic selection of more reliable modality information at each time step and channel level.

\subsubsection{Public Decoder}
Given fused memory representation $\mathbf{M}$, decoder aims to predict beam index classification results $\hat{y}_\tau$ for next $P$ steps. First, a target query sequence $\mathbf{T}^{(0)} \in \mathbb{R}^{P \times d}$ with length $P$ is constructed by initializing all queries as zero vectors and adding positional encoding, which is expressed as:
\begin{equation}
    \mathbf{t}_{\tau}^{(0)}
=
\mathbf{0}
+
\mathbf{p}_{\tau},
\quad
\tau = 1, \ldots, P,
\end{equation}
\begin{equation}
    \mathbf{T}^{(0)}
=
\left[
\mathbf{t}_1^{(0)},\,
\ldots,\,
\mathbf{t}_P^{(0)}
\right]^{\top}.
\end{equation}

Then, $L$ Transformer decoder layers are stacked. Layer $l$ is unfolded into three steps: self-attention, cross-attention, and feed-forward network, which are respectively expressed as:
\begin{equation}
    \left\{ \begin{array}{l}
{\bf{A}}_1^{(l)} = {\mathop{\rm LN}\nolimits} \left( {{{\bf{T}}^{(l - 1)}} + {\mathop{\rm MHA}\nolimits} \left( {{{\bf{T}}^{(l - 1)}},{\kern 1pt} {{\bf{T}}^{(l - 1)}},{\kern 1pt} {{\bf{T}}^{(l - 1)}}} \right)} \right)\\
{\bf{A}}_2^{(l)} = {\mathop{\rm LN}\nolimits} \left( {{\bf{A}}_1^{(l)} + {\mathop{\rm MHA}\nolimits} \left( {{\bf{A}}_1^{(l)},{\kern 1pt} {\bf{M}},{\kern 1pt} {\bf{M}}} \right)} \right)\\
{{\bf{T}}^{(l)}} = {\mathop{\rm LN}\nolimits} \left( {{\bf{A}}_2^{(l)} + {\mathop{\rm FFN}\nolimits} \left( {{\bf{A}}_2^{(l)}} \right)} \right),\quad l = 1, \ldots ,L.
\end{array} \right..
\end{equation}
Final decoder output is obtained as:
\begin{equation}
    \mathbf{T}^{(L)}
=
\left[
\mathbf{t}_1^{(L)},\,
\ldots,\,
\mathbf{t}_P^{(L)}
\right]^{\top}
\in
\mathbb{R}^{P \times d}.
\end{equation}

Finally, in beam prediction stage, hidden representation at each prediction step $\tau$ is mapped to $K$-class through a classification head, which is expressed as:
\begin{equation}
   \mathbf{o}_{\tau}
=
\mathbf{W}_o \mathbf{t}_{\tau}^{(L)}
+
\mathbf{b}_o
\in
\mathbb{R}^{K}, 
\end{equation}
and corresponding prediction probabilities and final beam index prediction results are respectively expressed as:
\begin{equation}
    \hat{\mathbf{p}}_{\tau}
=
\operatorname{softmax}
\left(
\mathbf{o}_{\tau}
\right),
\end{equation}
\begin{equation}
    \hat{y}_{\tau}
=
\arg\max_{k \in \{1, \ldots, K\}}
\hat{p}_{\tau,k}.
\end{equation}

\subsubsection{Training Process}
Algorithm~\ref{algor_hybrid} performs beam prediction by dynamically activating numeric, image, or multi-modal branches according to the inference mode. The selected features are temporally modelled and decoded to produce the predicted beam index sequence. Training of entire hybrid model system adopts sequence cross-entropy loss under supervised learning, and loss function is expressed as:
\begin{equation}
    \mathcal{L}=-
\sum_{\tau = 1}^{P}
\log
\hat{p}_{\tau, y_{\tau}} .
\end{equation}

\begin{figure}[!t]
  \centering
  \includegraphics[width=0.48\textwidth]{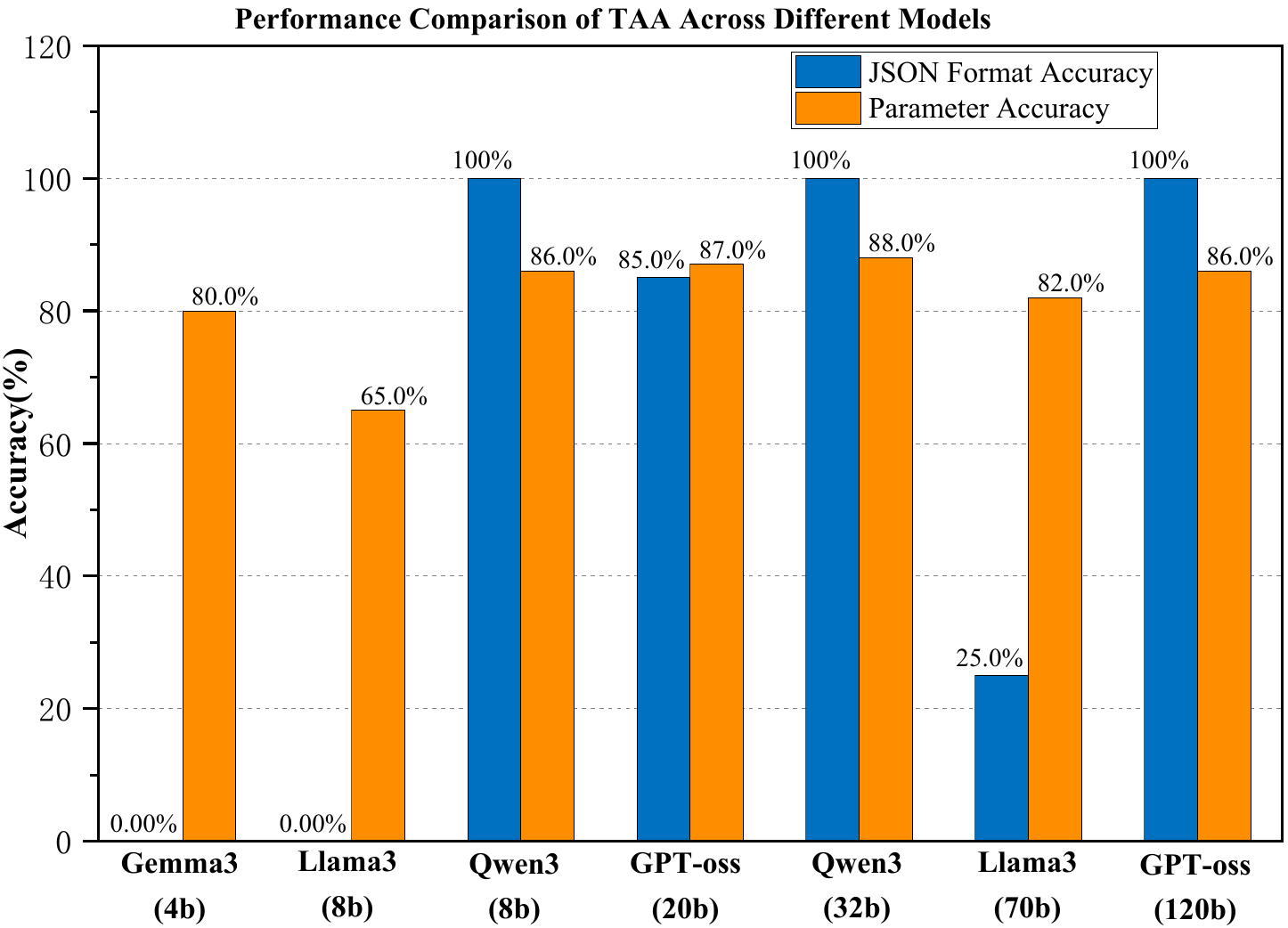}
  \caption{Performance of TAA in format accuracy and parameter accuracy.}
  \label{fig:TAA}
\end{figure}

\begin{figure*}[t]
\centering
\includegraphics[width=1.0\textwidth]{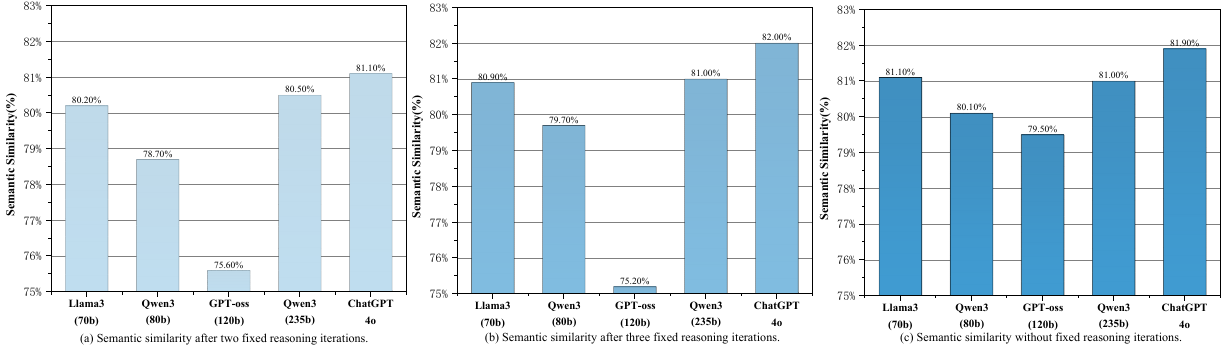} 
    \caption{Relationship between SPA reasoning iterations and reasoning results.}  
    \label{fig_SPA}         
\end{figure*}
\begin{figure}[t]
  \centering
  \includegraphics[width=0.48\textwidth]{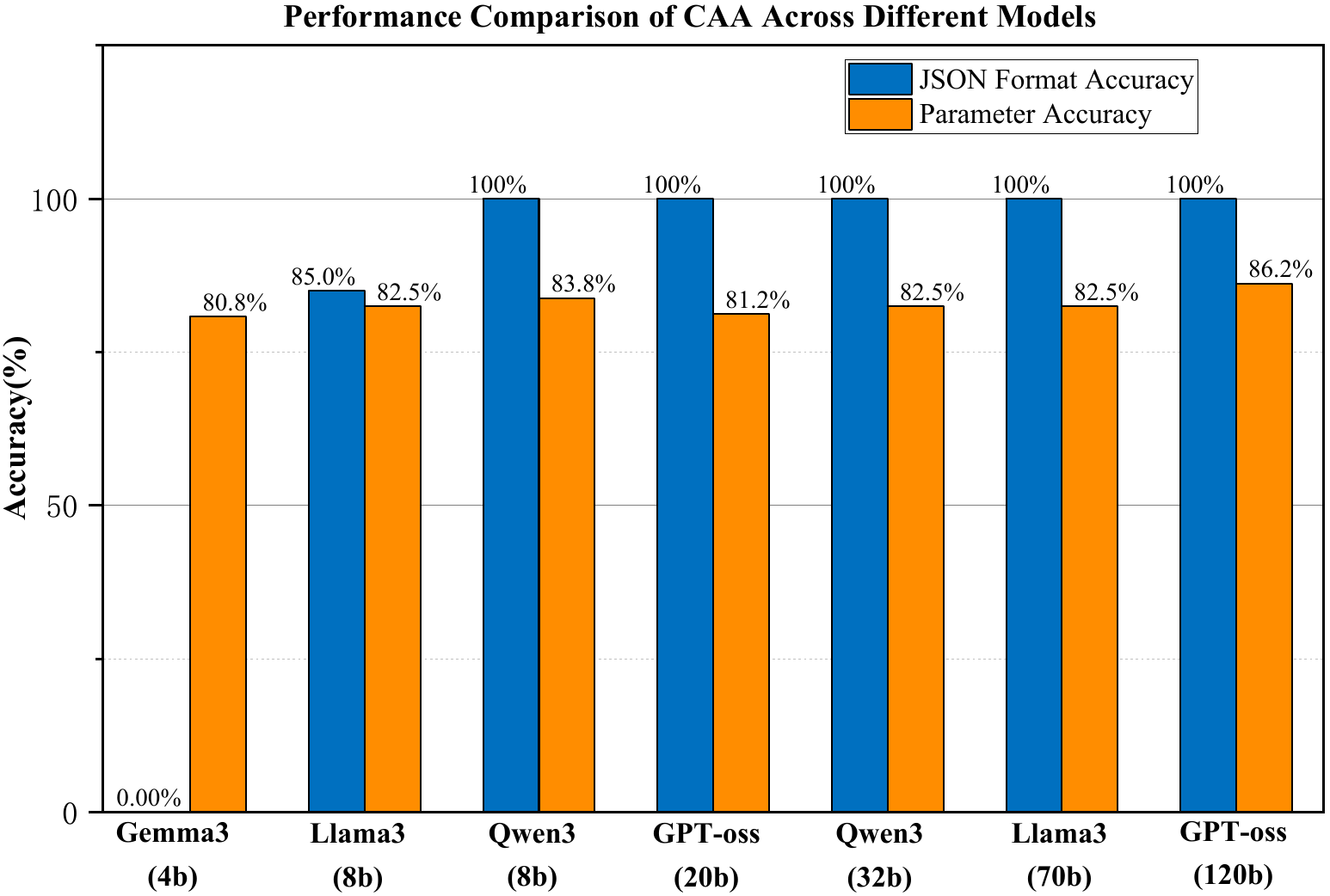}
  \caption{Performance of CAA in format accuracy and parameter accuracy.}
  \label{fig:CAA}
\end{figure}
\begin{figure*}[t]
\centering
\includegraphics[width=1.0\textwidth]{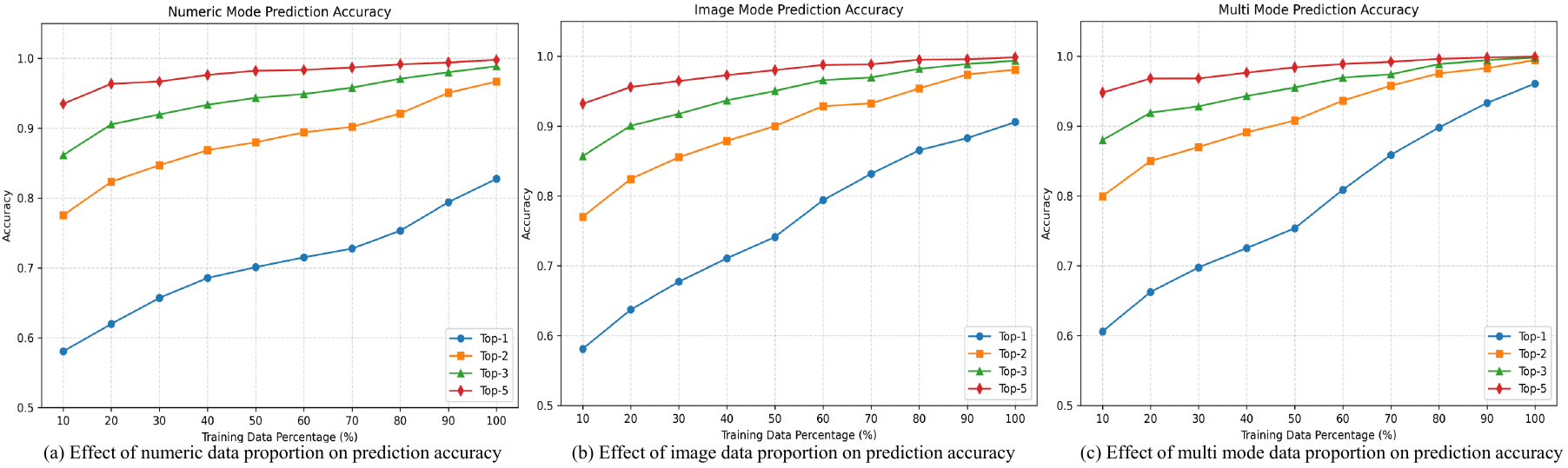} 
    \caption{Impact of training data proportion on beam prediction accuracy.}  
    \label{fig_training}         
\end{figure*}

\begin{figure*}[t]
\centering
\includegraphics[width=1.0\textwidth]{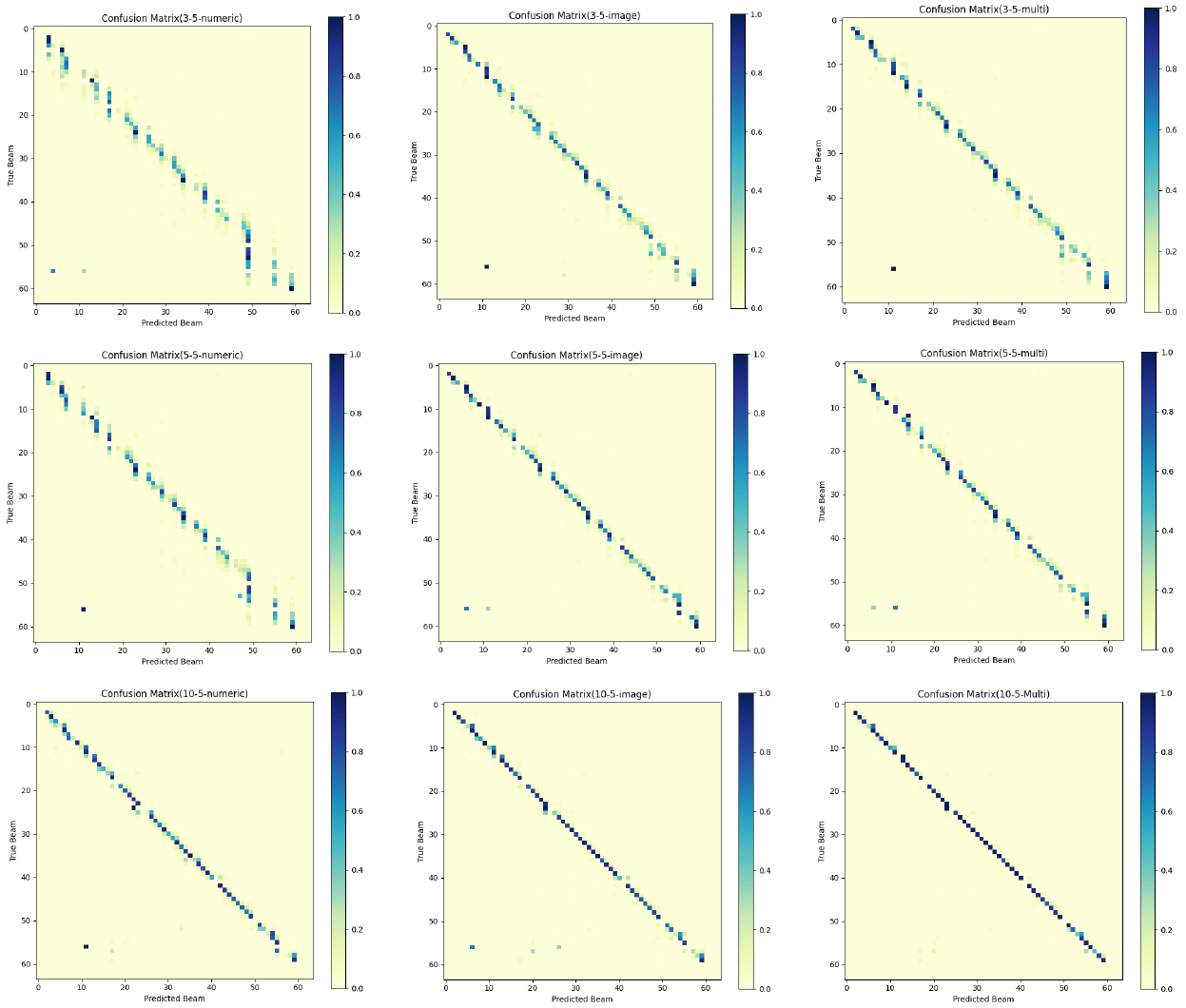} 
    \caption{Confusion matrices under different prediction scenarios.}  
    \label{fig_matrix}         
\end{figure*}

\section{Performance Evaluation}\label{V}
In this section, we design simulation experiments to evaluate effectiveness of multi-agent architecture and hybrid model system. We first describe dataset and simulation environment, and then present experimental results and analysis.

\subsection{Experimental Parameters}
We use a real UAV-to-mmWave base station dataset collected at Tuen Park in Arizona, named DeepSense6G \cite{10144504,9512383}. In this dataset, both base station and UAV are deployed in real environments. The base station is equipped with an RGB camera and a mmWave phased array, employing a 16-element phased array operating in the 60 GHz band and an oversampled codebook with 64 predefined beams for signal transmission and reception. The UAV is equipped with a mmWave transmitter, a GPS receiver, and an inertial measurement unit. \footnote{https://www.deepsense6g.net/scenarios/Scenarios\%2020-29/scenario-23}

\begin{table}[t]
\centering
\caption{Software Environment and Version Requirements}
\label{tab:software_versions}
\small
\setlength{\tabcolsep}{8pt}
\renewcommand{\arraystretch}{1.1}
\begin{tabular}{l c}
\hline
\textbf{Software / Library} & \textbf{Version} \\
\hline
Python        & 3.9 \\
PyTorch       & $\geq$ 2.0.0 \\
TorchVision   & $\geq$ 0.15.0 \\
Mamba-SSM     & $\geq$ 1.2.0 \\
NumPy         & $\geq$ 1.23 \\
Pandas        & $\geq$ 1.5 \\
Pillow        & $\geq$ 9.0 \\
TQDM          & $\geq$ 4.64 \\
\hline
\end{tabular}
\end{table}
Table \ref{tab:software_versions} lists software versions used in simulation experiments. CPU configuration is a 10-core Intel Xeon Processor (Skylake, IBRS), and GPU configuration is an NVIDIA A100-PCIE-40GB. In multi-agent experiments, we build interactive architecture based on LangChain and conduct comparative analysis using multiple open-source LLMs.
\subsection{Evaluation of Multi-Agent Architecture}
Fig.~\ref{fig:TAA} presents test results of TAA implemented using different LLMs, where horizontal axis denotes LLM type and vertical axis indicates accuracy percentage. TAA is responsible for extracting key elements from manager requirements and rewriting requirements in concise language. We construct 20 question sets with corresponding reference answers and evaluate accuracy based on TAA outputs. TAA outputs results in JSON format, including a rewritten requirement description, five key parameter elements (number of UAVs, data storage location, data format, data labels, and prediction accuracy requirement), and a flag indicating whether any element is missing. We sequentially input 20 questions into TAA and record 20 corresponding outputs. We then evaluate JSON format accuracy and accuracy of extracting five key parameter elements for these outputs. Experimental results show that both format accuracy and parameter accuracy improve as model size of adopted LLM increases. Among evaluated models, Qwen3-32B achieves best performance, while Llama3 exhibits weakest results, indicating that Qwen3 is better aligned with functionality of TAA.

Fig.~\ref{fig_SPA} presents evaluation results of SPA reasoning performance. SPA is responsible for analysing and reasoning over preprocessed requirements and producing solution plans. Similarly, we prepare 20 task sets preprocessed by TAA with corresponding reference plans and compare semantic similarity between SPA outputs and reference plans \cite{zhang2019bertscore}. We treat each think–act–observe cycle in the ReAct paradigm as one reasoning iteration and conduct comparative experiments with forced two iterations, forced three iterations, and unconstrained iteration numbers. Results are shown in Fig.~\ref{fig_SPA}(a), Fig.~\ref{fig_SPA}(b), and Fig.~\ref{fig_SPA}(c), respectively, where SPA without fixed iteration numbers uses a semantic similarity threshold of 75\%. Results indicate that SPA benefits from LLMs with larger parameter scales and that semantic similarity achieved with three reasoning iterations is generally higher than that with two iterations. Moreover, SPA can autonomously determine the number of reasoning iterations based on semantic similarity requirements and typically reaches a 75\% semantic similarity threshold within four iterations.

Fig.~\ref{fig:CAA} presents test results of CAA implemented using different LLMs, where horizontal axis denotes LLM type and vertical axis indicates accuracy percentage. CAA is responsible for verifying whether plan generated by SPA satisfies task requirements specified by TAA. Similar to TAA, we construct 20 sets of different inputs with predefined reference answers. Based on CAA responses, format accuracy and parameter accuracy are evaluated separately. CAA output includes a natural language part and a structured part. Natural language part describes task and solution plan, while structured part includes judgment on whether solution plan is successful and key parameters of prediction model, including model architecture, model parameters, and model inputs. Experimental results show that GPT-oss-120B is best aligned with CAA tasks, while Qwen3-8B also achieves strong performance. CAA tasks can be effectively executed on edge computing devices using Qwen3-8B for inference.

\subsection{Evaluation of Hybrid Beam Prediction Model System}

\begin{table*}[t]
\centering
\caption{Performance Comparison under Different Encoder Configurations}
\label{tab:encoder_comparison}
\small
\setlength{\tabcolsep}{6pt}
\renewcommand{\arraystretch}{1.2}
\begin{tabular}{lcccccc}
\toprule
\multicolumn{7}{c}{\textbf{Numeric Encoder Layer=1, Image Encoder Layer=1, Cross-Attention Head=1}} \\
\midrule
\textbf{Mode} & \textbf{Loss} & \textbf{top-1} & \textbf{top-2} & \textbf{top-3} & \textbf{top-5} & \textbf{Average per-frame inference latency (ms)} \\
\midrule
Numeric Mode & 0.7280 & 0.7600 & 0.9212 & 0.9634 & 0.9893 & 0.100 \\
Image Mode   & 0.5081 & 0.8152 & 0.9534 & 0.9804 & 0.9926 & 0.492 \\
Multi Mode   & 0.2979 & 0.8874 & 0.9787 & 0.9925 & 0.9980 & 0.550 \\
\midrule
\multicolumn{7}{c}{\textbf{Numeric Encoder Layer=2, Image Encoder Layer=2, Cross-Attention Head=2}} \\
\midrule
\textbf{Mode} & \textbf{Loss} & \textbf{top-1} & \textbf{top-2} & \textbf{top-3} & \textbf{top-5} & \textbf{Average per-frame inference latency (ms)} \\
\midrule
Numeric Mode & 0.4560 & 0.8276 & 0.9667 & 0.9888 & 0.9978 & 0.131 \\
Image Mode   & 0.2638 & 0.9058 & 0.9810 & 0.9935 & 0.9984 & 0.502 \\
Multi Mode   & 0.1188 & 0.9609 & 0.9945 & 0.9980 & 0.9994 & 0.597 \\
\midrule
\multicolumn{7}{c}{\textbf{Numeric Encoder Layer=4, Image Encoder Layer=4, Cross-Attention Head=1}} \\
\midrule
\textbf{Mode} & \textbf{Loss} & \textbf{top-1} & \textbf{top-2} & \textbf{top-3} & \textbf{top-5} & \textbf{Average per-frame inference latency (ms)} \\
\midrule
Numeric Mode & 0.4511 & 0.8411 & 0.9759 & 0.9902 & 0.9981 & 0.208 \\
Image Mode   & 0.2598 & 0.9177 & 0.9835 & 0.9940 & 0.9986 & 0.542 \\
Multi Mode   & 0.1237 & 0.9520 & 0.9941 & 0.9978 & 0.9993 & 0.697 \\
\midrule
\multicolumn{7}{c}{\textbf{Numeric Encoder Layer=4, Image Encoder Layer=4, Cross-Attention Head=2}} \\
\midrule
\textbf{Mode} & \textbf{Loss} & \textbf{top-1} & \textbf{top-2} & \textbf{top-3} & \textbf{top-5} & \textbf{Average per-frame inference latency (ms)} \\
\midrule
Numeric Mode & 0.4543 & 0.8401 & 0.9755 & 0.9901 & 0.9981 & 0.204 \\
Image Mode   & 0.2582 & \textbf{\textcolor{red}{0.9181}} & 0.9836 & 0.9941 & 0.9986 & 0.528 \\
Multi Mode   & 0.1003 & \textbf{\textcolor{red}{0.9657}} & 0.9956 & 0.9982 & 0.9994 & 0.687 \\
\midrule
\multicolumn{7}{c}{\textbf{Numeric Encoder Layer=4, Image Encoder Layer=4, Cross-Attention Head=4}} \\
\midrule
\textbf{Mode} & \textbf{Loss} & \textbf{top-1} & \textbf{top-2} & \textbf{top-3} & \textbf{top-5} & \textbf{Average per-frame inference latency (ms)} \\
\midrule
Numeric Mode & 0.4487 & \textbf{\textcolor{red}{0.8413}} & 0.9760 & 0.9902 & 0.9981 & 0.203 \\
Image Mode   & 0.2712 & 0.9123 & 0.9831 & 0.9938 & 0.9986 & 0.567 \\
Multi Mode   & 0.1299 & 0.9483 & 0.9938 & 0.9976 & 0.9992 & 0.700 \\
\bottomrule
\end{tabular}
\end{table*}
We first test parameter configurations of numeric encoder, image encoder, and multi-modal encoder, and results are listed in Table \ref{tab:encoder_comparison}. We use UAV data from past 10 seconds to predict beam index for next 5 seconds and adopt Top-$K$ accuracy to measure prediction performance, where correctness is defined by whether ground-truth labels appear among top $K$ predicted probabilities. Total number of inference samples is 3456, and average per-frame latency over all samples is also reported in the table. The numeric encoder consists of a linear projection layer followed by stacked Mamba residual blocks to model long-range temporal dependencies in numerical features. Number of Mamba residual blocks is treated as a hyper-parameter of numeric encoder and is set to 1, 2, and 4 layers, respectively. The image encoder adopts the same architecture as the numeric encoder, consisting of a ResNet backbone followed by stacked Mamba residual blocks. Number of Mamba residual blocks is treated as a hyper-parameter of image encoder and is set to 1, 2, and 4 layers, respectively. By employing multi-head cross-attention, the proposed multi-modal encoder allows numerical and visual representations to interact through multiple complementary subspaces, thereby enhancing the robustness of cross-modal alignment under dynamic UAV mobility. Therefore, we treat the number of cross-attention heads as a hyper-parameter of the multi-modal encoder and set it to 1, 2, and 4, respectively. 

Table \ref{tab:encoder_comparison} systematically compares model performance and inference overhead under different encoder configurations across three operating modes: numerical modality (Numeric), image modality (Image), and multimodal modality (Multi). Results show that as numbers of layers in numeric and image encoders increase from 1 to 4, overall model performance improves significantly. Among all configurations, multi mode consistently achieves lowest loss and highest top-1 to top-5 accuracy, validating effectiveness of fusing numeric and visual information. In addition, increasing number of cross-attention heads from 1 to 2 or 4 further improves top-1 accuracy to some extent. This effect is most evident under 4-layer encoder settings, where multi model reaches highest top-1 accuracy, although performance gains exhibit diminishing returns. For numeric, image, and multi modes, highest top-1 accuracy reaches 0.8413, 0.9181, and 0.9657, respectively. Multi-view modelling of cross-modality relationships explains why multi mode achieves higher accuracy than single-modality modes. For numeric or image data, hybrid model system models long-range dependencies within each modality using Mamba, and 4 cross-attention heads are sufficient to capture information such as position, velocity, distance, and historical beams. Further increasing number of heads leads to parameter growth and noise amplification. In contrast, inference latency increases steadily with encoder depth and attention head count. Multi mode consistently incurs higher computational overhead than single-modality modes, reflecting trade-off between performance improvement and computational complexity. During beam prediction plan formulation, SPA derives corresponding solution plans through in-depth understanding of task requirements and aforementioned encoders.

Fig.~\ref{fig_training} illustrates impact of training data quantity on beam prediction accuracy. We curate 11387 samples from DeepSense6G, among which 8064 samples are used for training and 3456 samples are used for testing. In Fig.~\ref{fig_training}, a training data ratio of 100\% indicates that all 8064 training samples are used. We continue to use past 10 seconds of data to predict beam index for future 5 seconds as testing scenario. It can be observed that prediction accuracy increases steadily as number of training samples grows, indicating that sufficient training data enable the model to better learn channel characteristics and environmental dynamics, thereby improving reliability of beam selection. When training data reach a certain scale, accuracy gains gradually diminish and performance tends to saturate, suggesting that model approaches its representational capacity limit under current architecture and parameter configuration.

Fig.~\ref{fig_matrix} presents comparative confusion matrices of numeric, image, and multimodal models under different prediction scenarios. Prediction scenarios include using past 3 seconds, 5 seconds, and 10 seconds of data to predict beam index for future 5 seconds. It can be observed that prediction results of all three modalities exhibit a clear diagonal-dominant distribution, indicating that models can correctly predict corresponding beam indices in most cases. Compared with single-modality models, multimodal model shows more concentrated diagonal energy and significantly reduced off-diagonal elements, especially when past 5 seconds or 10 seconds of data are available, where misclassification is further suppressed, demonstrating stronger robustness and stability. In addition, as amount of past data increases, confusion matrices gradually approach an ideal diagonal structure, indicating that looser prediction constraints effectively improve beam hit probability and validating advantage of multimodal fusion in complex beam prediction tasks.

\section{Conclusion}\label{vI}
This paper investigated embodied beam prediction for low-altitude UAV communications and proposed a multi-agent agentic AI architecture combined with a hybrid multimodal beam prediction model system. The proposed architecture decomposed beam prediction into structured reasoning stages through task analysis, solution planning, and completeness assessment, effectively addressing the limitations of single-LLM-based approaches. A hybrid beam prediction model was designed to jointly exploit numeric mobility data and visual information using Mamba-based temporal modelling and cross-attention-based multimodal fusion, with data-flow strategies adaptively determined by inter-agent reasoning. Simulation results on a real-world UAV dataset demonstrated that the proposed approach achieved superior beam prediction accuracy, robustness, and adaptability compared with single-modality and conventional methods. The results confirmed that integrating agentic AI with multimodal deep learning provides an effective and scalable solution for proactive beam prediction in highly dynamic low-altitude economy networks.

In our future research, we plan to introduce continual learning to enhance beam prediction accuracy of the hybrid model system under varying data distributions and feature patterns. We also consider integrating UAV trajectory optimization into the proposed multi-agent reasoning architecture.

\bibliographystyle{IEEEtran}
\bibliography{bibRef}

\end{document}